\documentclass[aoas,MSNbibl,nameyear,dvips]{arximspdf}
\usepackage{breakurl}

\doi{10.1214/10-AOAS398F}
\referstodoi{10.1214/10-AOAS398}
\volume{5}
\issue{1}
\pubyear{2011}
\firstpage{88}
\lastpage{90}



\begin{document}
\begin{frontmatter}

\title{Discussion of:  A statistical analysis of multiple temperature proxies: Are
reconstructions of surface temperatures over the last 1000 years reliable?}
\runtitle{Discussion}
\pdftitle{Discussion on A statistical analysis of multiple temperature proxies:
Are reconstructions of surface temperatures over the last 1000 years reliable?
 by B. B. McShane and A. J. Wyner}

\begin{aug}
\author[A]{\fnms{Peter} \snm{Craigmile}\ead[label=e1]{pfc@stat.osu.edu}\thanksref{t1}}
\and
\author[B]{\fnms{Bala} \snm{Rajaratnam}\corref{}\ead[label=e2]{brajarat@stanford.edu}\thanksref{t2}}

\runauthor{P. Craigmile and B. Rajaratnam}

\affiliation{The Ohio State University and Stanford University}

\address[A]{Department of Statistics\\
The Ohio State University\\
Columbus, Ohio 43210\\
USA\\
\printead{e1}} 

\address[B]{Department of Statistics\\
Department of Environmental Earth \\
System Science\\
The Woods Institute for the Environment\\
Stanford University\\
Stanford, California 94305\\
USA\\
\printead{e2}}
\end{aug}
\thankstext{t1}{Supported in part by NSF Grants DMS-06-04963 and DMS-09-06864.}
\thankstext{t2}{Supported in part by NSF Grants DMS-09-06392,
AGS-1003823 and Grants SU-WI-EVP10, SUFSC08-SUFSC10-SMSCVISG0906.}

\received{\smonth{9} \syear{2010}}


%

\end{frontmatter}

Professors McShane and Wyner have written a thought-provoking paper
that intends to challenge some of the conventional wisdom in the paleoclimate
literature. Rather than commenting on the merits of the entire
methodology we focus on one topic. Namely, we discuss
theoretical and practical aspects of the use of the least absolute
shrinkage and selection operator [\citet{Tibshirani1996}], more
popularly known as the ``Lasso,'' in the context of paleoclimate reconstruction.

It is important to acknowledge at first sight that the Lasso seems like
a natural candidate in the paleoclimate context, since one is
immediately faced with a larger number of proxies, compared to the
number of data points [e.g., in \citet{McShaneWyner2010} (hereafter
MW), Section 3.2, the response variable is of length 149 whereas there
are 1138 predictors]. It is clear that standard regression-based variable
selection techniques will not work. The sheer number of predictors
does indeed warrant a need for
regularization. Many techniques are available for such problems,
including popular methods such as ridge regression and principal
component regression.

As pointed out by MW the ``Lasso tends to choose sparse $\widehat
{\beta}^{\,\mathrm{Lasso}}$ thus serving as a variable selection
methodology and alleviating the $p\gg n$ problem.'' This point is very
well taken. The model selection capability of the Lasso has made it
very relevant in this era of high throughput data and rapidly changing
information technology. Consequently the Lasso has been useful in
biomedical and genomic applications where genes are often in the tens
of thousands, compared to much fewer subjects. Biomedical scientists
often wish to isolate a few, but important genes that are related to
disease conditions. The Lasso ``zeroes out'' smaller coefficients and
thus can be used for model selection.

In a more abstract setting, consider a statistical model such as a
regression model which has a low signal-to-noise ratio where the
coefficient vector is not sparse. It is quite easy to see that the
Lasso can exclude many predictors which have small but nonzero
coefficients. This exclusion will occur with a higher degree of
severity, a feature that is not available in ridge regression or
principal components regression. The Lasso works well if the signal is
sparse; that is, there are few large nonzero coefficients and many
true zero coefficients. Just as with many other estimators, the signal
needs to be also bounded away from zero for the Lasso to be able to
recover the nonzero coefficients accurately.
Thus, it is not clear immediately whether the very model selection
feature that makes the Lasso attractive in so many settings is as
equally desirable in the paleoclimate context. As the authors state
correctly, the relationship between the predictors and response
variable is weak. Hence the coefficient estimates are very small in
magnitude. In
this instance the Lasso could potentially zero out many of those
coefficients. It is quite feasible that the proxies could collectively
have some predictive power though the contribution of each of the
individual proxy time-series may be rather small [\citet{Lietal2007}].

There are also other reasons why the Lasso may not always be
appropriate in the paleoclimate context. First, the Lasso can choose at
most $n$ nonzero coefficients [see \citet{Efronetal2004}]. Hence, by
design, any other additional set of proxies which may have \textit{almost}
the same predictive power, but slightly less than the first $n$
predictors, will have no chance of being selected in the model. This
problem comes back to our original point that in the paleoclimate
context there may be more than a few sparse signals, but rather a large
number of weak signals instead. This is simply due to the irregular,
dependent noise in the data and the structural relationships between
instrumental records and paleoclimate proxies [see, e.g., \citet
{Tingleyetal2010}]. Furthermore, the standard Lasso does not yield any
ridge or
Steinian-type shrinkage, a feature that can potentially lead to better
RMSE. In this regard,
the elastic net proposed by \citet{ZouHastie2005} might be more
suitable. Second, while the Lasso has the capability to do model
selection by zeroing out certain variables, it also has the adverse
effect of shrinking even the larger nonzero coefficients via
soft-thresholding. This indiscriminate shrinking of the coefficients
leads to biased estimates. Third, the Lasso is not an oracle procedure,
and there are scenarios in which Lasso variable selection is
inconsistent [\citet{Zou2006}]. Such theoretical safeguards, under
broad assumptions, could be very useful, especially in a hotly debated
topic such as climate change.
\citet{Zou2006} proposes the adaptive Lasso as a possible remedy
for these problems and may be worthy of exploration in this context.

A further issue involves the fact that ``proxy series
contain very complicated and highly autocorrelated time series
structures'' (MW). The standard Lasso assumes that the errors in the
regression model are uncorrelated, which is clearly not the case
here. Indeed, a review of statistical models appearing in
Section 5.2 of MW points to significant autocorrelations that have to be
accounted for. Further research is needed, especially for
paleoclimatic variable selection [see \citet{GelperCroux2008} for a
time series version of the Lasso applied to economics data]. This
is further complicated by the ``problem of spatial correlation'' which
MW choose to ignore in their article.

In closing, we have indicated some of our concerns of using the Lasso
for paleoclimatic reconstruction. This does not mean that variable
selection is unimportant---it is. For example, with tree proxies an
argument can be made [e.g., \citet{Tingleyetal2010}, Section~3] that
only trees
in certain areas contain climate signatures. In defense of
MW we note that they do consider techniques other than the
Lasso in Section 4 (space limitations exclude us from commenting on
these methods). We conclude by commending the authors
on a thought-provoking paper, and by referring the reader to a
recent manuscript [\citet{Tingleyetal2010}] that sheds further light
and gives
detailed statistical insights into some of the important issues in
paleoclimatic reconstruction.


\printaddresses

\end{document}